\documentclass[preprint,aps, showpacs, showkeys]{revtex4}

\usepackage{graphicx}

\begin{document}

\title{Modulational instability in asymmetric coupled wave functions
}

\author{
I. Kourakis\footnote{Electronic address:
\texttt{ioannis@tp4.rub.de} ;
\texttt{www.tp4.rub.de/$\sim$ioannis}.} and P.K.
Shukla\footnote{Electronic address: \texttt{ps@tp4.rub.de} ;
\texttt{www.tp4.rub.de/$\sim$ps}.}}\affiliation{Institut f\"ur
Theoretische Physik IV,
Fakult\"at f\"ur Physik und Astronomie, \\
Ruhr-Universit\"at Bochum, D-44780 Bochum, Germany}

\date{Submitted 13 October 2005}

\begin{abstract}
The evolution of the amplitude of two nonlinearly interacting
waves is considered, via a set of coupled nonlinear
Schr\"odinger-type equations. The dynamical profile is determined
by the wave dispersion laws (i.e. the group velocities and the GVD
terms) and the nonlinearity and coupling coefficients, on which no
assumption is made. A generalized dispersion relation is obtained,
relating the frequency and wave-number of a small perturbation
around a coupled monochromatic (Stokes') wave solution. Explicitly
stability criteria are obtained. The analysis reveals a number of
possibilities. Two (individually) stable systems may be
destabilized due to coupling. Unstable systems may, when coupled,
present an enhanced instability growth rate, for an extended wave
number range of values. Distinct unstable wavenumber windows may
arise simultaneously.
\end{abstract}



\maketitle

Amplitude modulation (AM) is a widely known nonlinear mechanism
dominating wave propagation in dispersive media \cite{Dauxois}; it
is related to mechanisms such as modulational instability (MI),
harmonic generation and energy localization, and possibly leads to
soliton formation. The study of AM generically relies on nonlinear
Schr\"odinger (NLS) type equations \cite{Sulem}; a set of coupled
NLS (CNLS) equations naturally occurs when interacting modulated
waves are considered. CNLS equations are encountered in physical
contexts as diverse as electromagnetic wave propagation
\cite{Ostrovskii,MKB}, optical fibers \cite{Hasegawa3,Agrawal},
plasma waves \cite{Das, Gupta,IKNJP}, transmission lines
\cite{Bilbault}, and left-handed (negative refraction index)
metamaterials (LHM) \cite{Lazarides}. A similar mathematical model
is employed in the mean-field statistical mechanical description
of boson gases, to model the dynamics of Bose-Einstein condensates
\cite{Dalfovo, Kasamatsu, EPJB}. In this paper, we shall
investigate the (conditions for the) occurrence of MI in a pair of
(asymmetric) CNLS equations, from first principles. A set of
stability criteria are derived, to be tailor cut to a (any)
particular problem of coupled wave propagation.

\paragraph{The model.}

Let us consider two coupled waves propagating in a dispersive
nonlinear medium. The wave functions ($j= 1, 2$) are modelled by
$\psi_j \exp i(\mathbf{k_j r} - \omega_j t) + {\rm c.c.}$ (complex
conjugate), where the carrier wave number $\mathbf{k_j}$ and
frequency $\omega_j$ of each wave are related by a dispersion
relation function $\omega_j = \omega_j(\mathbf{k_j})$.
Nonlinearity is manifested via a slow modulation of the wave
amplitudes, in time and space, say along the $x-$axis. The
amplitude evolution is described by a pair of CNLS Eqs.
\begin{eqnarray}
i \biggl( \frac{\partial \psi_1}{\partial t} + v_{g, 1}
\frac{\partial \psi_1}{\partial x} \biggr) + P_1 \frac{\partial^2
\psi_1}{\partial x^2} +  Q_{11} \,|\psi_1|^2  \psi_1 +
Q_{12} \,|\psi_2|^2  \psi_1 = 0 \, , \nonumber \\
i \biggl( \frac{\partial \psi_2}{\partial t} + v_{g, 2}
\frac{\partial \psi_2}{\partial x} \biggr) + P_2 \frac{\partial^2
\psi_2}{\partial x^2} +  Q_{22} \,|\psi_2|^2  \psi_2 + Q_{21}
\,|\psi_1|^2  \psi_2 = 0 \, . \label{CNLSE}
\end{eqnarray}
The group velocity $v_{g,j}$ and the group-velocity-dispersion
(GVD) term $P_j$ corresponding to the $j-$th wave is related to
(the slope and the curvature, respectively, of) the dispersion
curve via $v_{g,j} = \partial \omega_j/\partial k_x$ and $P_{j} =
\partial^2 \omega_j/2\partial k_x^2$ (differentiation \emph{in the
direction of modulation}). $Q_{jj}$ and $Q_{jj'}$ model carrier
self-modulation and wave coupling, respectively. No hypothesis
holds, \emph{a priori}, on the sign and/or the magnitude of either
of these coefficients. The group velocities are often assumed
equal, in which case (and only) the corresponding terms are
readily eliminated via a Galilean transformation. The combined
assumption $P_1 = P_2$, $Q_{11} = Q_{22}$ and $Q_{12} = Q_{21}$ is
often made in nonlinear optics \cite{MKB, Lisak}. The case $P_1 =
P_2$, $Q_{11} = Q_{21}$ and $Q_{12} = Q_{22}$ was also recently
reported, in negative refarction index composite metamaterials
\cite{Lazarides}.

\paragraph{Modulational (in)stability of single waves.\label{1wave}}

Let us first briefly outline, for later reference, the analysis in
the case of a single modulated wave, say here recovered by setting
$\psi_2 = 0$ in Eqs. (\ref{CNLSE}); the (single) NLS equation is
thus obtained. According to the standard formalism \cite{Dauxois,
Sulem}, $\psi_1$ ($=\psi$, dropping the index in this paragraph)
is \emph{modulationally unstable} (stable) if $P Q > 0$ ($P Q <
0$). To see this, one may first check that the NLSE is satisfied
by the plane wave solution
 \( \psi(x, t) = \psi_0\, e^{i Q |\psi_0|^2 t} \).
The standard (linear) stability analysis then shows that a linear
perturbation (say $\psi_0 \rightarrow \psi_0 + \epsilon
\delta\psi_0$) with frequency $\Omega$ and wavenumber $K$ [i.e.
$\delta\psi_0 \sim \exp i(K x - \Omega t)$] obeys the dispersion
relation:
\( (\Omega - v_{g} K)^2 = P \, K^2\, \bigl( P \,K^2\, - 2 Q
\,|\psi_0|^2 \bigr)\), which exhibits a purely growing unstable
mode if \(K \leq K_{cr, 0} = (2 Q/P)^{1/2}\,|\psi_0|\) (hence only
if $P Q > 0$). The growth rate $\sigma = \rm{Im}\Omega$ attains a
maximum value \( \sigma_{max} = Q \,|\psi_0|^2 \) at $K_{cr,
0}/\sqrt{2}$. For $P Q < 0$, on the other hand, the wave is
\emph{stable} to external perturbations.

\paragraph{Coupled wave stability analysis.\label{MI}}

In order to investigate the modulational stability profile of a
pair of coupled waves, we shall first seek an equilibrium state in
the form $\psi_j = \psi_{j0} \exp[i\varphi_j (t)]$ (for $j=1,2$),
where $\psi_{j0}$ is a (constant real) amplitude and $\varphi_j
(t)$ is a (real) phase, into Eqs. (\ref{CNLSE}). We thus find a
monochromatic (fixed-frequency) solution of the form $\varphi_j
(t) = \Omega_{j0} t$, where \(\Omega_{j0} = Q_{jj} \psi_{j0}^2 +
Q_{jl} \psi_{l0}^2 \) (for $j \neq l=1, 2$, henceforth understood
everywhere). Considering a small perturbation around equilibrium,
we take $\psi_j =(\psi_{j0} + \epsilon \psi_{j1})\exp[i\varphi_j
(t)]$, where $\psi_{j1}({\bf r}, t)$ is a small ($\epsilon \ll 1$)
complex amplitude perturbation of the wave amplitudes.
Substituting into Eqs. (\ref{CNLSE}) and separating real and
imaginary parts by writing $\psi_{j1} =a_j + i b_j$ (where $a_j,
b_j \in \Re$), the first order terms (in $\epsilon$) yield
\begin{eqnarray}
- \frac{\partial b_j}{\partial t} - v_{g,j} \frac{\partial
b_j}{\partial x} + P_j \frac{\partial^2 a_j}{\partial x^2} + 2
Q_{jj}
\psi_{j0}^2 a_j + 2 Q_{jl} \psi_{j0} \psi_{l0} a_l =0 \, , \nonumber \\
\frac{\partial a_j}{\partial t}  + v_{g,j} \frac{\partial
a_j}{\partial x}+  P_j \frac{\partial^2 b_j}{\partial x^2} = 0 \,
. \label{perteqs1}
\end{eqnarray}
Eliminating $b_j$, these equations yield
\begin{eqnarray}
\biggl[ \biggl( \frac{\partial }{\partial t} + v_{g,j}
\frac{\partial }{\partial x} \biggr)^2 + P_1 \biggl(P_1
\frac{\partial^2 }{\partial x^2} + 2 Q_{11} \psi_{10}^2\biggr)
\frac{\partial^2 }{\partial x^2} \biggr] a_1 + P_1 Q_{12}
\psi_{10} \psi_{20} \frac{\partial^2 }{\partial x^2} a_2 = 0 \, ,
\quad \label{perteqs2}
\end{eqnarray}
(along with a symmetric equation, upon $1 \leftrightarrow 2$). Let
$a_j = a_{j0} \exp[i(K x - \Omega t)] + $ c.c., where $K$ and
$\Omega$ are the wavevector and the frequency of the perturbation,
respectively, viz. $\partial/\partial t \rightarrow - i \Omega$
and $\partial/\partial x \rightarrow i K$.
We  thus obtain an eigenvalue problem in the form $\mathbf{M a} =
\Omega^2 \mathbf{a}$, where $\mathbf{a} = (a_{10}, a_{20})^T$ and
the matrix elements are $M_{jj} = P_j K^2 \, (P_j K^2 - 2 Q_{jj}
\psi_{j0}^2) \equiv \Omega_{j}^2$ and $M_{jl} = - 2 P_j Q_{jl}
\psi_{j0} \psi_{l0} K^2$ (for $l \ne j = 1 $ or $2$)
\begin{equation}
\bigl[ (\Omega - v_{g, 1} K)^2 - \Omega_1^2 \bigr] \bigl[ (\Omega
- v_{g, 2} K)^2 - \Omega_2^2 \bigr] = \Omega_{c}^4 \,  \label{DR0}
\end{equation}
where $\Omega_c^4 = M_{12} M_{21}$. This dispersion relation is a
4th order polynomial equation in $\Omega$.

\paragraph{Equal group velocities.}

For $v_{g, 1} = v_{g, 2}$, setting $\Omega - v_{g, 1/2}
K\rightarrow \Omega$ reduces (\ref{DR0}) to
\begin{equation}
\Omega^4 - T \Omega^2 + D = 0 \, , \label{DR2}
\end{equation}
where $T = {\rm{Tr}}\mathbf{M} \equiv \Omega_{1}^2 + \Omega_{2}^2$
and $D = {\rm{Det}}\mathbf{M} \equiv \Omega_{1}^2 \Omega_{2}^2 -
\Omega_c^4$ are the \textsl{trace} and the \textsl{determinant},
respectively, of the matrix $\mathbf{M}$. Eq. (\ref{DR2}) admits
the solution \( \Omega^2 =\frac {1}{2} \bigl[ T \pm (T^2 -4
D)^{1/2} \bigr]\), or
\begin{equation}
\Omega_{\pm}^2 =\frac {1}{2} (\Omega_1^2+\Omega_2^2) \pm \frac
{1}{2} \left[ (\Omega_1^2 - \Omega_2^2)^2 + 4
\Omega_{c}^4\right]^{1/2} \, . \label{solution2}
\end{equation}

Stability is ensured (for any wavenumber $K$) if (and only if)
\emph{both} of the (two) solutions of (\ref{DR2}), say
$\Omega_{\pm}^2$, are \emph{positive} (real) numbers. In order for
the right-hand side to be real, the \textsl{discriminant} quantity
$\Delta = T^2 - 4 D = (\Omega_1^2 - \Omega_2^2)^2 + 4 \Omega_c^4$
has to be positive. Furthermore, recalling that the roots of the
polynomial $p(x) = x^2 -T x + D$, say $r=r_{1,2}$, satisfy $T =
r_1 + r_2$ and $D = r_1 r_2$, the stability requirement is
tantamount to the following three conditions being satisfied
simultaneously: $T > 0$, $D > 0$ \emph{and} $\Delta = T^2 - 4 D
> 0$.

The first stability condition, namely the positivity of the trace
$T$: \(T = K^2 [K^2 \sum_j P_j^2 - 2 \sum_j P_j Q_{jj}
\psi_{j0}^2] \,
> \, 0
\), depends on (the sign of) the quantity $q_1 \equiv \sum_j P_j
Q_{jj} |\psi_{j0}|^2$ which has to be negative for stability. The
only case ensuring \emph{absolute} stability (for any $\psi_{j0}$
and $k$) is
\begin{equation}
P_1 Q_{11}< 0 \quad and \quad  P_2 Q_{22} < 0 \, . \label{C1}
\end{equation}
Otherwise, $T$ becomes negative (and thus either $\Omega_-^2 < 0 <
\Omega_+^2$ or $\Omega_-^2 < \Omega_+^2 < 0$) for $K$ below a
critical value $K_{cr, 1} = (2 \sum_j P_j Q_{jj}
\psi_{j0}^2/\sum_j P_j^2)^{1/2}
> 0$ (cf. the single wave criterion above); this is always possible
for a sufficiently large perturbation amplitude $|\psi_{20}|$ if,
say, $P_2 Q_{22} > 0$ (even if $P_1 Q_{11} < 0$). Therefore, only
a pair of two individually stable waves can be stable, or the
presence of a \emph{single} unstable wave may de-stabilize its
counterpart.

The second stability condition, namely positivity of the
determinant $D$, amounts to
\[D(K^2) = P_1 P_2 K^4 \bigl[ (P_1 K^2 - 2 Q_{11} \psi_{10}^2) \,
(P_2 K^2 - 2 Q_{22} \psi_{20}^2) - 4 Q_{12} Q_{21} \psi_{10}^2
\psi_{20}^2 \bigr] \,  > \, 0 \, .
\]
We see that $D(K^2)$ bears two non-zero roots for $K^2$, namely
$K_{D,1/D,2}^2 = \frac{1}{2 P_1^2 P_2^2} [q_2 \mp (q_2^2 - 4 P_1^2
P_2^2 q_3)^{1/2}]$, as obvious from the expanded form $D = k^4
(P_1 P_2 K^4 - q_2 K^2 + q_3)$, where $q_2 \equiv \sum K_{D,j}^2 =
2 P_1 P_2 (P_2 Q_{11} \psi_{0, 1}^2 + P_1 Q_{22} \psi_{0, 2}^2)$
and $q_3 \equiv \prod K_{D,j}^2 = 4 P_1 P_2 (Q_{11} Q_{22} -
Q_{12} Q_{21}) \psi_{0, 1}^2 \psi_{0, 2}^2$. The condition $D
> 0 \quad (\forall K \in \Re)$
requires either:\\
-- that the discriminant quantity $\Delta' = q_2^2 - 4 P_1^2 P_2^2
q_3 = 4 P_1^2 P_2^2 \, [(P_1 Q_{22} \psi_{20}^2 - P_2 Q_{11}
\psi_{10}^2)^2 + 4 P_1 P_2 Q_{12} Q_{21} \psi_{10}^2 \psi_{20}^2]$
be non-positive, i.e. $\Delta' \leq 0$ (this is only possible if
$P_1 P_2 Q_{12} Q_{21} < 0$ \emph{and} for a specific relation to
be satisfied by the perturbation amplitudes $\psi_{j0}$; that is,
it \emph{cannot} be generally satisfied, $\forall \psi_{j0}$), or
\\
-- that $\Delta' > 0$ and \emph{both} of the (real) non-zero roots
$K_{D,1/D,2}^2$ of $D(K^2)$ be \emph{negative}; this is ensured if
$q_2 \,  < \, 0 $ \emph{and} $q_3  > 0$. If $q_3 > 0$ and $ q_2
> 0$, then the two roots $K_{D,1/2}^2$ will be positive ($0 <
K_{D,1}^2 < K_{D,2}^2$) and the wave pair will be unstable to a
perturbation with intermediate $K$, i.e. $K_{D,1}^2 < K^2 <
K_{D,2}^2$. If $q_3 < 0$ (regardless of $q_2$), then $K_{D,1}^2 <
0 < K_{D,2}^2$, and the wave pair is unstable to a perturbation
with $K^2 < K_{D,2}^2$.
\\ These
instability scenaria and wavenumber thresholds are sufficient for
symmetric wave systems (i.e. upon $1 \leftrightarrow 2$), as we
shall see below.

The last stability condition regards the positivity of the
discriminant quantity $\Delta = T^2 - 4 D$ (irrelevant if $D <
0$). We consider the inequality
\[\Delta(k^2) =
K^4 \, (d_4 K^4 - d_2 K^2 + d_0) \, > \, 0 \,
\]
where $d_4 = (P_1^2 - P_2^2)^2$, $d_2 = 4 (P_1^2 - P_2^2) (P_1
Q_{11} \psi_{10}^2 - P_2 Q_{22} \psi_{20}^2)$ and $d_0 = 4[(P_1
Q_{11} \psi_{10}^2 - P_2 Q_{22} \psi_{20}^2)^2 + 4 P_1 P_2 Q_{12}
Q_{21} \psi_{10}^2 \psi_{20}^2]$. We should distinguish two cases
here.

If $P_1 = P_2= P$, this condition reduces to $d_0 > 0$, i.e. here
$d'_0 = 4 P^2 [(Q_{11} \psi_{10}^2 - Q_{22} \psi_{20}^2)^2 + 4
Q_{12} Q_{21} \psi_{10}^2 \psi_{20}^2] \equiv 4 P^2 q_4$.
Stability (for all $\psi_{j0}$) is thus only ensured if $Q_{12}
Q_{21} > 0$. For symmetric wave pairs, i.e. for $P_1 = P_2= P$ and
$Q_{12} = Q_{21}$, this last necessary condition for stability is
always fulfilled. If, on the other hand, $Q_{12} Q_{21} \le 0$,
the wave pair will be unstable in a range of values (e.g. of the
ratio $\psi_{10}/\psi_{20}$), to be determined by solving $d'_0 <
0$).

Let us now assume (with no loss of generality) that $P_1 > P_2$.
Since $\Delta'' = d_2^2 - 4 d_4 d_0 = - 64 P_1 P_2 Q_{12} Q_{21}
(P_1^2 - P_2^2)^2 \psi_{10}^2 \psi_{20}^2$, the stability
condition $\Delta
> 0$ is satisfied \emph{for all} $K$ and $\psi_{j0}$ only if the quantity
$q_5 \equiv P_1 P_2 Q_{12} Q_{21}$ is positive, hence $\Delta'' <
0$; again, this is always true for symmetric waves. Now if, on the
other hand, $q_5 < 0$ (i.e. $\Delta''
> 0$), then one needs to investigate the signs of
$d_2 = K_{\Delta, 1}^2 + K_{\Delta, 2}^2 \equiv q_6$ and $d_0 =
K_{\Delta, 1}^2 K_{\Delta, 2}^2 \equiv q_7$, in terms of the
amplitudes $\psi_{j0}$. Here, $K_{\Delta, 1/2}^2 = \frac{1}{2 d_4}
[d_2 \mp \sqrt{d_2^2 - 4 d_4 d_0}]$. Similar to the analysis of
the previous condition (see above), one may easily see that both
signs are possible for both quantities $d_2$ and $d_0$. The only
possibility for stability ($\forall K$) is provided by the
combination $d_2 < 0$ and $d_0 > 0$ (hence $K_{\Delta, 1}^2 <
K_{\Delta, 2}^2 < 0$). The possibility for instability arises
either for $K_{\Delta, 1}^2  < 0 < K^2 < K_{\Delta, 2}^2$ (if $d_0
< 0$), or for $0 < K_{\Delta, 1}^2 < K^2 < K_{\Delta, 2}^2$ (if
$d_0 > 0$ \emph{and} $d_2 > 0$). As above, see that we obtain the
possibility for a window of instability far from $K=0$.

Instability is manifested as a purely growing mode, when one or
more of the above conditions are violated. In specific, if $T < 0$
and/or $D < 0$, then one (or both) of the solutions of the
dispersion relation (\ref{DR0}) (for $\Omega^2$) becomes negative,
say $\Omega_-^2 < 0$ [given by (\ref{solution2})]; the instability
growth rate in this case is given by $\sigma \equiv \sqrt{-
\Omega_-^2}$,  and is manifested in the wavenumber ranges $[0,
K_{cr, 1}]$ and either $[0, K_{D, 2}]$ or $[K_{D, 1}, K_{D, 2}]$
(depending on parameter values; see the definitions above).

If $\Delta = T^2 - 4 D < 0$, on the other hand (hence $D > 0$),
then all solutions of (\ref{DR0}) are complex, thus developing  an
imaginary part ${\rm Im}(\Omega_\pm^2) = \pm \sqrt{|\Delta}|/2$,
so (the maximum value of) ${\rm Im}(\Omega_\pm) = {\rm
Im}(\Omega_\pm^2)^{1/2}$ gives the instability growth rate
$\sigma$. As found above, this will be possible for wave numbers
either in $[0, K_{\Delta, 2}]$ or $[K_{\Delta, 1}, K_{\Delta, 2}]$
(see the definitions above).

\begin{figure}[htb]
 \centering
 \resizebox{2.1in}{!}{
 \includegraphics[]{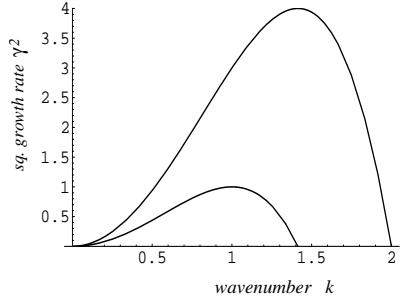}}
\caption{The square of the instability growth rate
$\gamma^2 \equiv -\Omega_-^2$ is depicted versus the perturbation
wavenumber $K$ (arbitrary parameter values).
Notice the difference from the single wave case (lower curve). }
\label{Kourakis2Fig1}
\end{figure}

The analysis indicates that up to three different unstable
wavenumber ``windows'' may appear; these windows may either be
partially superposed, or distinct from each other. One may
therefore qualitatively anticipate MI occurring for $K \in [0,
K_{cr}]$ (some threshold) and, \emph{also}, for $K \in [K'_{cr},
K''_{cr}]$ ($K'_{cr}$ may be higher or lower than $K_{cr}$,
depending on the problem's parameters). Furthermore, the instability growth rate witnessed may be dramatically modified by the coupling, both quantitatively (higher rate) and qualitatively (enlarged unstable wavenumber region); see in Fig. \ref{Kourakis2Fig1}.

Summarizing the above results, should one wish to investigate the
occurrence of modulational instability in a given physical
problem, one has to verify condition (\ref{C1}), and then consider
the (sign of the) quantities $q_1, ... q_7$ (defined above).

\paragraph{The role of the group velocity misfit.}

It may be interesting to discuss the role of the group velocity
difference, in a coupled wave system. Keeping the discussion
qualitative, we shall avoid to burden the presentation with
tedious numerical calculations. One may rather point out the role
of the group velocity misfit via simple geometric arguments.
Inspired by an idea proposed in Ref. \cite{Das}, we may express
the general dispersion relation (\ref{DR0}) in the form
\begin{equation}
f_1(x) = f_2(x) \label{DR0bis}
\end{equation}
where we have defined the functions \( f_1(x) = (x - x_1)^2 + A\)
and \(f_2(x) = \frac{C}{(x - x_2)^2 + B} \), and the real
quantities $x_j =  K v_{g, j}$, $A = - \Omega_1^2 = - M_{11}$, $B
= - \Omega_2^2 = - M_{22}$, and $C = \Omega_{c}^4 = M_{12}
M_{21}$; $x$ here denotes $\Omega$. The stability profile is
determined by the number of \emph{real} solutions of Eq.
(\ref{DR0bis}), an integer, say $r$, between 0 and 4. For absolute
stability (for any $K$, $|{\cal E}_{j0}|$), we need to have 4 real
solutions; in any other case, i.e. if $r < 4$, the (imaginary part
of) the $4 - r$ complex solutions determine(s) the growth rate of
the instability. Note that $x_1 \ne x_2$ expresses the group
velocity mismatch $v_{g, 1} \ne v_{g, 2}$. Negative $A$ ($B$)
means that wave 1 (2) alone is stable, and vice versa.

\begin{figure}[htb]
 \centering
 \resizebox{2.1in}{!}{
 \includegraphics[]{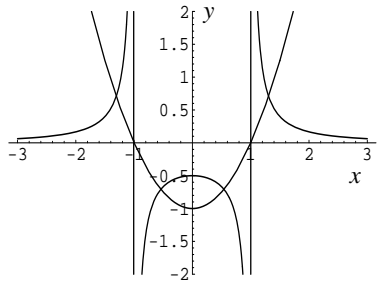}
} \caption{The functions $f_1(x)$ (parabola) and $f_2(x)$
(rational function, two vertical asymptotes) defined in the text
are depicted, vs. $x$, for $A = B = -1$, $C = 0.5$ (so that $D = A
B - C = +0.5 > 0$), $x_1 = x_2 = 0$ (equal group velocities). Note
that a group velocity mismatch (a horizontal shift) may
destabilize a pair of (stable, separately) waves (i.e. reduce the
intersection points from 4 to 2).} \label{Kourakis2Fig2}
\end{figure}

\begin{figure}[htb]
 \centering
 \resizebox{2in}{!}{
 \includegraphics[]{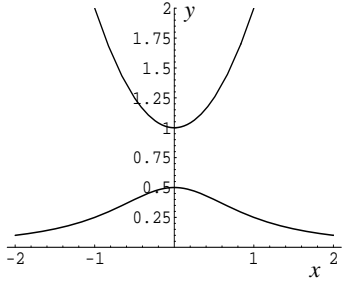}}
\caption{The functions $f_1(x)$ and $f_2(x)$ are depicted, for $A
= B = +1$, $C = 0.5$ (so that $D = AB - C > 0$), and $x_1 = x_2 =
0$. At most 2 intersection points may occur by translation. A pair
of (unstable, separately) waves is always unstable.}
\label{Kourakis2Fig3}
\end{figure}

Let us first consider a wave pair satisfying $C
> 0$, i.e. for $M_{12} M_{21} \sim P_1 P_2 Q_{12} Q_{21}> 0$
(a \emph{symmetric} wave pair, for instance). We shall study the
curves representing the functions $f_1(x)$ and $f_2(x)$ on the
$xy$ plane. The former one is a parabola, with a minimum at $(x_1,
A)$. The latter one is characterized by a local maximum (for $C
> 0$) at $(x_2, C/B)$,
in addition to a horizontal asymptote (the $x$-axis), since
$f_2(x) \rightarrow 0$ for $x \rightarrow \pm \infty$.
Furthermore, for $B < 0$ (only), $f_2(x)$ has two vertical
asymptotes (poles) at $x = x_2 \pm \sqrt{|B|}$ (see Figs.
\ref{Kourakis2Fig2}, \ref{Kourakis2Fig3}). Now, for a \emph{stable
- stable} wave pair (i.e. for $A, B < 0$), we have seen that the
dispersion relation (\ref{DR0}) predicted stability. This result
regarded the \emph{equal} (or vanishing) group velocity case,
$v_{g, 1} = v_{g, 2}$, and may be visualized by plotting $f_1(x)$
and  $f_2(x)$ for $x_1 = x_2$ and $A, B < 0$; see Fig.
\ref{Kourakis2Fig2}a. Let us first assume that $D = M_{11} M_{22}
- M_{12} M_{21} = A B - C$ is \emph{positive}, implying (for $B <
0$) that $A < C/B$; thus, the minimum of $f_1(x)$ lies below the
local maximum of $f_2(x)$. Thus, 4 points of intersection exist
(cf. Fig. \ref{Kourakis2Fig2}a), for $x_1 = x_2$ and $A, B < 0$;
this fact ensures stability, as we saw above via analytical
arguments, for $v_{g, 1} = v_{g, 2}$ and $M_{11}, M_{22}
> 0$ (both waves individually stable). Now, considering $v_{g, 1}
\ne v_{g, 2}$ results in a horizontal shift between the two curves
(cf. Fig. \ref{Kourakis2Fig2}b), which may exactly result in
reducing the number of intersection points from 4 to 2 (enabling
instability). Therefore, \emph{a pair of stable waves may be
destabilized due to a finite difference in group velocity}.

Still for a stable-stable wave pair ($A, B < 0$), let us assume
that $D = A B - C < 0$, implying (for $B < 0$) that $A > C/B$.
Thus, the minimum of $f_1(x)$ here lies \emph{above} the local
maximum of $f_2(x)$, and only 2 points of intersection now exist,
(shift the parabola upwards in Fig. \ref{Kourakis2Fig2} to see
this); this fact imposes \emph{instability} (for $D < 0$),
as predicted above.

Considering an \emph{unstable - unstable} wave pair (i.e.
 $A > 0$ and $B > 0$) with
$D = A B - C > 0$ ($A > C/B$). Plotting $f_1(x)$ and $f_2(x)$ for
$x_1 = x_2$ and $A, B > 0$ (see Fig. \ref{Kourakis2Fig3}), we see
that the minimum of $f_1(x)$ lies above the local maximum of
$f_2(x)$. \emph{No} points of intersection exist, a fact which
prescribes \emph{instability}. Considering $v_{g, 1} \ne v_{g, 2}$
simply results in a horizontal shift between the two curves, which
\emph{does not} affect this result at all. On the other hand
(still for $A > 0$ and $B > 0$), now assuming that $D = A B - C <
0$, i.e. $A < C/B$, results in a vertical shift downwards of the
parabola in Fig. \ref{Kourakis2Fig3}a; at least 2 complex
solutions obviously exist, hence instability. Therefore, \emph{a
pair of unstable waves is always unstable} ($\forall \quad v_{g,
1}, v_{g, 2}$).

Still for $C > 0$, one may consider a \emph{stable - unstable}
wave pair (say, for $A < 0$ and $B > 0$, with no loss of
generality): the plot of $f_{1}$ and $f_{2}$ (here omitted) would
look like Fig. \ref{Kourakis2Fig3} upon a strong vertical
translation of the parabola downwards (so that the minimum lies in
the lower half-plane, since $A < 0$). Instability ($r =2$)
dominates this case also.

Let us now consider a wave pair satisfying $C < 0$, i.e. $M_{12}
M_{21} \sim P_1 P_2 Q_{12} Q_{21} = P^2 Q^2 < 0$; this has to be
an \emph{asymmetric} wave pair. Again, different cases may be
distinguished.

\begin{figure}[htb]
 \centering
 \resizebox{2in}{!}{
 \includegraphics[]{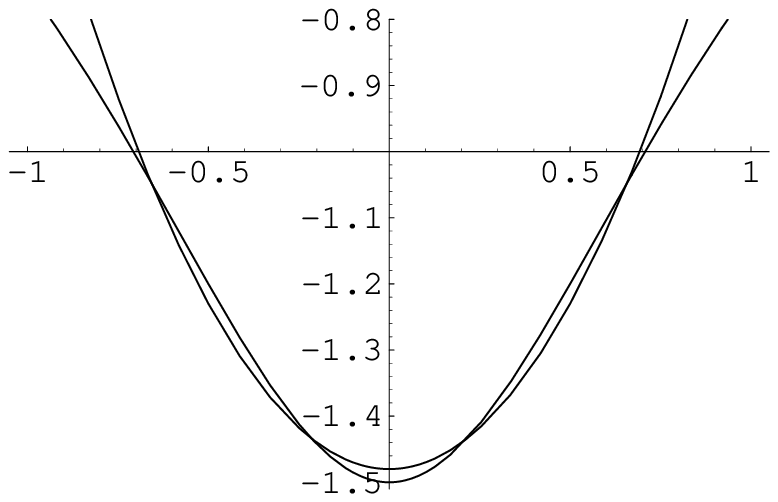}}
 \caption{\emph{Stable-unstable} wave interaction:
 the functions $f_1(x)$ and $f_2(x)$ are
depicted, for $A = -1.48$, $B = +1$, $C = -1.5$, and $x_1 = x_2 =
0$. This (stable, 4 intersection points) configuration may be
destabilized either by a horizontal ($v_g$ difference) or a
vertical ($A$ value) shift.} \label{Kourakis2Fig4}
\end{figure}

For an \emph{stable-unstable} wave pair, i.e. say for $A < 0 < B$,
different possibilities exist: cf. Fig. \ref{Kourakis2Fig4}, where
4 points of intersection ensure stability, for $v_{g, 1} \approx
v_{g, 2}$). However, either  a (horizontal) shift $v_g$ difference
or (a vertical shift) in $A$  may render the system unstable.

For $A, B > 0$ (both waves intrinsically \emph{unstable}), one
easily sees that no intersection occurs (figure omitted; simply
translate the parabola upwards in Fig. \ref{Kourakis2Fig4}); the
pair is unstable.

\begin{figure}[htb]
 \centering
 \resizebox{2in}{!}{
 \includegraphics[]{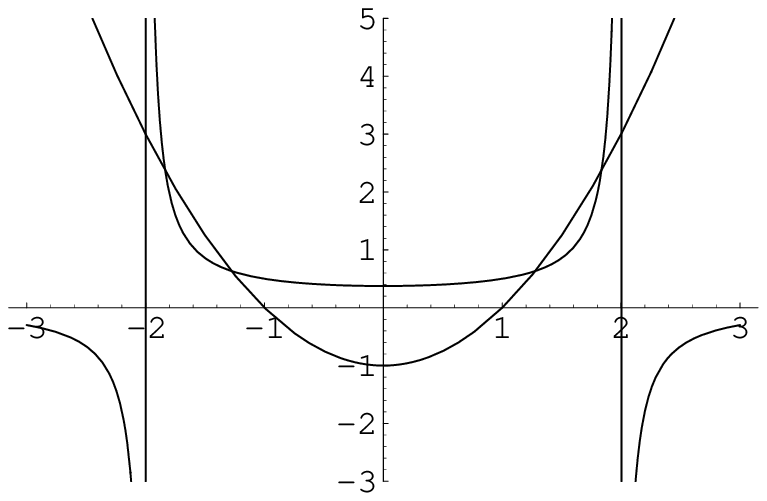}}
 \caption{\emph{Stable-stable} wave interaction:
 the functions $f_1(x)$ and $f_2(x)$ are
depicted, for $A = -1$, $B = -4$, $C = -1.5$ and $x_1 = x_2 = 0$.
This (stable, 4 intersection points) configuration may be
destabilized either by a horizontal ($v_g$ difference) or a
vertical ($A$ value) shift.} \label{Kourakis2Fig5}
\end{figure}

Finally, for a \emph{stable-stable} wave pair ($A, B < 0$), the
wave pair may stable (see Fig. \ref{Kourakis2Fig5}); this
configuration is nevertheless destabilized either by a velocity
misfit (a horizontal shift) or a vertical shift (in $A$).

\emph{In conclusion}, we have investigated the occurrence of
modulational instability in a pair of coupled waves,
co-propagating and interacting with one another. Relying on a
coupled NLS equation model, we have derived a complete set of
explicit (in)stability criteria, in addition to exact expressions
for the critical wavenumber thresholds. Furthermore, we have
traced the role of the group velocity mismatch on the coupled wave
stability. The results are readily applied to a set of coupled
Gross-Pitaevskii equations (modelling a pair of BECs in condensed
boson gases), as exposed here, as well as in a variety of physical
situations.




\bigskip

\textbf{Figures}


\newpage

\end{document}